\documentclass[10pt,letterpaper]{article}
\usepackage{opex3}
\usepackage{graphicx}
\usepackage{dcolumn}
\usepackage{amsmath}
\usepackage{epsfig}
\usepackage[]{subfigure}
\usepackage{rangecite}
\usepackage{color}

\def\opex{ Opt.\ Express }

\def\jqe{ IEEE J.\ Quantum Electron.\ }

\def\josab{ J.\ Opt.\ Soc.\ Am.\ B }

\def\pra{ Phys.\ Rev.\ A }
\def\prb{ Phys.\ Rev.\ B }

\def\prl{ Phys.\ Rev.\ Lett.\ }
\def\rmp{ Rev.\ Mod.\ Phys.\ }

\begin{document}




\title{A quantum-dynamical theory for nonlinear optical interactions in graphene}

\author{Zheshen Zhang$^{1,2}$ and Paul L. Voss$^{1,2}$}

\address{1. Georgia Tech Lorraine, Georgia Tech-C.N.R.S., UMI
2958, \\2-3 rue Marconi, Metz, France \\ and \\
2. School of Electrical and Computer Engineering, \\Georgia
Institute of Technology, \\
777 Atlantic Drive NW 30332-0250 Atlanta, USA}

\email{voss@ece.gatech.edu} 

\begin{abstract}
We use a quantum-dynamical model to investigate the optical response of graphene under low excitation power. Ultrafast carrier relaxation processes, which play an important role for understanding the optical response of graphene, are included phenomenologically into the model. We obtain analytical solutions for the linear and third-order nonlinear optical response of graphene, and four-wave mixing in particular. This theory shows agreement with recently reported experimental data on linear complex optical conductivity and four-wave mixing, providing evidence for ultrafast quantum-dephasing times of approximately 1 fs.
\end{abstract}

\ocis{(190.4380)Nonlinear optics, four-wave mixing; (270.5565) Quantum Communications;} 


\section{Introduction}
Graphene is a two-dimensional material consisting of a monolayer of carbon atom forming hexagonal lattices \cite{neto09}. Much has been learned since its first isolation from graphite and deposition onto a dielectric substrate \cite{novoselov04}. Due to its linear and massless band structure near the Dirac points, graphene has been shown to posses astonishing linear electronic and optical properties. Electron mobility in graphene reaches $2\times10^5$ cm$^2$/Vs, much higher than normal materials such as silicon \cite{du08}. Graphene is also the only known material exhibiting anomalous quantum Hall effect under room temperature \cite{gusynin06}. The optical absorption per layer of graphene is related to the fine-structure constant by $\pi\alpha = 2.3\%$ over a broad range of terahertz and optical wavelengths \cite{nair08,stauber08,mecklenburg10}. Potential applications include ultrabroadband fast detectors, and the replacement of transparent conductive oxides with graphene. Also of interest are potential coherent electronics applications that would use quantum interference of electrons to provide new device functionality.

Besides linear electronic and optical properties, the third-order nonlinear response of materials is useful for certain applications such as wavelength conversion \cite{harter80} and quantum-entanglement generation \cite{li05}. The most important physical parameter for nonlinear-optical experiments is the nonlinear coefficient, typically the nonlinear susceptibility, of the material used. However, in practice, nonlinear-optical experiments are limited by other factors, in particular, the phase-matching conditions arising from wavelength-dependent refractive indices. The accumulation of phase-mismatch as the interacting waves propagate through the nonlinear medium sets a limit on the bandwidths of the nonlinear applications. Because the graphene thickness is much less than an optical wavelength, non-critical phase-matching occurs simply, though at the cost of some optical absorption.

Nonlinear properties of graphene have been investigated theoretically in \cite{mikhailov07,mikhailov08,rosenstein10,dora10,ishikawa10}. Theoretical results predict high nonlinearity of graphene at the terahertz-frequency \cite{mikhailov07,mikhailov08} and optical-frequency \cite{ishikawa10} ranges. More recently, the nonlinearity of graphene in the visible-optical range was experimentally measured by four-wave mixing \cite{hendry10}. When taking into account the nonlinear response per unit thickness, the theory adopted in Ref. \cite{hendry10} shows that graphene's effective third-order nonlinear susceptibility $\chi^{(3)}$  is of $8$ orders higher than glass. Ref. \cite{hendry10} estimated the $ \chi^{(3)} $ of graphene by evaluating the nonlinear surface-current density induced by an external electromagnetic field and estimating the dynamic nonlinear response of electrons coupled with an external field is characterized by the electron-photon interaction Hamiltonian. Instead of the dipole interaction Hamiltonian utilized by \cite{hendry10}, a more general form of the interaction Hamiltonian can be obtained by the minimal substitution $\textbf{p}\rightarrow\textbf{p}-q\textbf{A}(t)$, giving a more accurate description of the dynamics. As we will see in this paper, the dipole interaction Hamiltonian is not adequate to describe the quantum dynamics of electrons in graphene. Additionally, the quantum dynamics of electrons in graphene is also affected by ultrafast many-body interactions such as electron-electron and electron-phonon scattering \cite{sun08,haug04,piscanec04}, which in turn may depend on carrier densities, conduction band energies, and temperature in non-trivial ways, resulting in ultrafast quantum dephasing. Other quantum dephasing mechanisms may exist. A complete physical model needs to include these effects, which have not yet been included in previous theoretical calculations of the nonlinearity of graphene.

In this paper, we present a quantum-dynamical model for investigating the quantum dynamics of electrons in graphene. The electron-photon interaction Hamiltonian is obtained by the minimal substitution on the free-electron Hamiltonian to relate the quantum dynamics of electrons with the vector potential of the electromagnetic field, as in \cite{rosenstein10}. This paper differs in that it includes electron-electron and electron-phonon scattering by introducing two phenomenological decay rates. This paper also focuses on nonlinear mixing rather than on the linear decay-free, dephasing-free model. In a quasi-continuous-wave pump experiment, in which the electron-photon coupling time (order of ps) is much longer than the carrier-relaxation time (order of fs), these phenomenological decay rates allow us to avoid dealing with complicated microscopic quantum-mechanical calculations and to obtain analytical solutions for quantum dynamics of electrons in graphene by exploiting quantum-perturbation method. The proposed model is appropriate for excitation power well below the saturation threshold of graphene (4 GW/cm$^2$) \cite{xing10}, in which case the population inversion for each quantum state in the Brillioun zone does not significantly change. Once the quantum dynamics of electrons in graphene is obtained, both linear and nonlinear optical conductivity produced by different optical processes will be derived.

This paper is organized as follows. In Sec. 2, we discuss the quantum dynamics of electrons in graphene. Our discussion in this section covers the electron-photon interaction and ultrafast quantum dephasing caused by carrier-relaxation processes in graphene. In Sec. 3, we derive the surface-current density in graphene. We calculate both the linear and nonlinear surface-current densities that are caused by the driving optical frequency fields. We also discuss the linear complex optical conductivity and connect it to ultrafast quantum dephasing in graphene. In Sec. 4, we apply the nonlinear surface-current density obtained in Sec. 3 to evaluate four-wave mixing in graphene, and fit the theory to the recently reported experimental data \cite{hendry10}. In Sec. 5, we discuss the validity of the proposed quautum-dynamical model under certain physical circumstances. Conclusions are reviewed in Sec. 6.

\section{Quantum dynamics of electrons in graphene}
In this section, the quantum dynamics of electrons in graphene will be reviewed, starting from first principles. We begin with the free-electron Hamiltonian, obtained for the 2D tight binding model of nearest neighbour interaction approximation. The electron-photon interaction Hamiltonian is derived later by minimal substitution. We then discuss the ultrafast carrier relaxation, which plays an important role in the quantum dynamics of electrons in graphene. The resulted ultrafast quantum dephasing in graphene is included by addition of phenomenological decay rates into the model.

\subsection{Free-electron Hamiltonian and electron-photon coupling}
In the 2D tight-binding model, or nearest neighbour interaction approximation, the motion of electrons is limited by assuming that they can only hop to their nearest neighbours. The Hamiltonian of electrons in graphene under this assumption is written in the second-quantized language \cite{neto09}:
\begin{equation}
\label{eqHamiltonianSpace}
\hat{H} = -\eta\sum_{\langle i,j\rangle,\sigma}(a^\dag_{\sigma,i}b_{\sigma,j}+H.c.),
\end{equation}
where $\eta\approx2.8 eV$ is the hopping energy and the product of the creation and annihilation operators $a^\dag_{\sigma,i}b_{\sigma,j}$ illustrates a process in which an electron with spin $\sigma \in \{\uparrow,\downarrow\}$ is annihilated on site $R_i$ of sublattice A and a new electron of the same spin is created on its neighbour site $R_j$ of sublattice B. The Fourier transform of the Hamiltonian in Eq. \ref{eqHamiltonianSpace} allows the derivation of the free-electron Hamiltonian in the momentum space, which is
\begin{equation}
\label{eqHamiltonian_k}
\hat{H} = \sum_{\textbf{k}}(a^\dag_\textbf{k}b^\dag_{\textbf{k}})H_0(^{a_\textbf{k}}_{b_\textbf{k}}),
\end{equation}
where
\begin{equation}
H_0=\left(
\begin{array}{cc}
0&h_{\textbf{p}}\\
h^*_{\textbf{p}}&0
\end{array}
\right)
\end{equation}
is the first-quantized free-electron Hamiltonian, characterizing the quantum dynamics of a single electron with momentum $\textbf{p} = \hbar \textbf{k}$. The summation in Eq. \ref{eqHamiltonian_k} is performed over the entire Brillioun zone. In the vicinities of two Dirac points $ \textbf{K}^\pm $ in the Brillioun zone, we have the following linear dispersion relation for $ h_{\textbf{p}} $:

\begin{equation}
\label{eqhplinear}
h_{\textbf{p}} = v_F(p_x \mp ip_y) = v_Fp(\cos\theta \mp i\sin\theta),
\end{equation}
where $v_F \approx 10^6$ m/s is the Fermi velocity of electrons in graphene. Energy eigenstates and corresponding eigen-energies can be found by solving the 2D Dirac equation \cite{kao10}:
\begin{equation}
H_0 |\varphi\rangle = E|\varphi\rangle.
\end{equation}
Given the electron momentum $ \textbf{p} $, one finds two energy eigenstates with opposite eigenvalues $E_{C(V)} = \pm |h_{\textbf{p}}| = \pm v_F p$, corresponding to the conduction-band and valence-band respectively:
\begin{eqnarray}
|C_\textbf{p}\rangle&=&\frac{1}{\sqrt{2}}\left(^1_{h^*_\textbf{p}/|h_\textbf{p}|}\right)\\
|V_\textbf{p}\rangle&=&\frac{1}{\sqrt{2}}\left(^1_{-h^*_\textbf{p}/|h_\textbf{p}|}\right).
\end{eqnarray}

Having discussed the Hamiltonian for a single free electron in graphene, we next derive Hamiltonian of an electron coupled to an external electromagnetic field, which is characterized by its vector potential $\textbf{A}(t)$, related to the electric field $\textbf{E}$ and the magnetic field $\textbf{B}$ by
\begin{eqnarray}
\label{eqVectorPotentialtoFields}
\textbf{E} &=& -\frac{\partial \textbf{A}}{\partial t}\\\notag
\textbf{B} &=& \nabla\times\textbf{A}.
\end{eqnarray}

Suppose at time $t = 0$, we turn on a homogeneous a.c. electromagnetic field having a vector potential with two polarization components:
\begin{equation}
\widetilde{\textbf{A}}(t) = \widetilde{A}_x(t)\hat{\textbf{x}}+\widetilde{A}_y(t)\hat{\textbf{y}} = \left(\widetilde{A}_x(t),\widetilde{A}_y(t)\right),
\end{equation}
where $\hat{\textbf{x}}$ and $\hat{\textbf{y}}$ are two orthogonal unit vectors of the Cartesian coordinate. Each polarization component $\widetilde{A}_q(t)$, $q\in\{x,y\}$, is the summation of different frequency modes on a polarization:
\begin{eqnarray}
\widetilde{A}_q(t) = \sum_{n}A_{q,n}e^{-i\omega_{q,n}t} = \sum_n \widetilde{A}_{q,n}(t).
\end{eqnarray}
Both positive and negative frequencies are allowed. The complex conjugate relation $A_{q,n} = A^*_{q,-n}$ holds to guarantee a real field. By using Eq. \ref{eqVectorPotentialtoFields}, the electric field on polarization $ q $ is related to the vector potential by
\begin{equation}
\widetilde{E}_q(t) = -\frac{d\widetilde{A}_q(t)}{dt} = \sum_n i\omega_{q,n}A_{q,n}e^{-i\omega_{q,n}t} = \sum_n E_{q,n}e^{-i\omega_{q,n}t}.
\end{equation}
The electric-field amplitude and the vector-potential amplitude have the following relation:
\begin{equation}
\label{eqEARelation}
E_{q,n}=i\omega_{q,n}A_{q,n}.
\end{equation}

The electron-photon interaction Hamiltonian can be obtained by the minimal substitution $\textbf{p}\rightarrow \textbf{p}+e\widetilde{\textbf{A}}(t)$ on the free-electron Hamiltonian $H_0$ \cite{shankar94}. The Taylor expansion of the electron-photon interaction Hamiltonian around $\textbf{p}$ reads
\begin{equation}
\label{eqGrapheneHamiltonian}
H(t)=\left(
\begin{array}{cc}
0&h_\textbf{p}+\nabla h_\textbf{p}\cdot e\widetilde{\textbf{A}}(t)\\
h^*_\textbf{p}+\nabla h^*_\textbf{p} \cdot e\widetilde{\textbf{A}}(t)&0
\end{array}
\right).
\end{equation}
Due to the linear band structure of graphene, higher-order terms in the Taylor expansion vanish in the vicinities of two Dirac points. Let $H(t) = H_0 + V(t)$, so that we have
\begin{equation}
V(t) = \left(
\begin{array}{cc}
0&\nabla h_\textbf{p}\cdot e\widetilde{\textbf{A}}(t)\\
\nabla h^*_\textbf{p} \cdot e\widetilde{\textbf{A}}(t)&0
\end{array}
\right)
= e \sum_u \widetilde{A}_u(t)\left(
\begin{array}{cc}
0&\frac{\partial h_\textbf{p}}{\partial p_u}\\
\frac{\partial h^*_\textbf{p}}{\partial p_u} &0
\end{array}
\right),
\end{equation}
where $ u\in \{x,y\} $. In the representation of energy eigenstates $|V_\textbf{p}\rangle$ and $|C_\textbf{p}\rangle$, $V(t)$ is organized into the following matrix form:
\begin{eqnarray}
\label{eqVtMatrix}
V(t) = \left(
\begin{array}{cc}
\langle V_\textbf{p}|V(t)|V_\textbf{p}\rangle & \langle V_\textbf{p}|V(t)|C_\textbf{p}\rangle\\
\langle C_\textbf{p}|V(t)|V_\textbf{p}\rangle&\langle C_\textbf{p}|V(t)|C_\textbf{p}\rangle
\end{array}
\right)
&=&\sum_{u,j} e \widetilde{A}_{u,j}(t)\left(
\begin{array}{cc}
V_{VV}^{u}& V_{VC}^{u}\\
V_{CV}^{u}&V_{CC}^{u}
\end{array}
\right).
\end{eqnarray}
Around the two Dirac points $ \textbf{K}^\pm $, by use of Eq. \ref{eqhplinear}, one obtains
\begin{eqnarray}
V_{VV}^x &=& -v_F\cos\theta, V_{VC}^x = \pm iv_F\sin\theta, V_{CV}^x = \mp iv_F\sin\theta, V_{CC}^x = v_F\cos\theta\notag\\
V_{VV}^y &=& -v_F\sin\theta, V_{VC}^y =  \mp iv_F\cos\theta, V_{CV}^y = \pm iv_F\cos\theta, V_{CC}^y = v_F\sin\theta.
\end{eqnarray}

In general, the quantum dynamics of an electron can be obtained by solving the time-dependent Schr\"{o}dinger equation:
\begin{equation}
\label{eqSchrodingerMotion}
i\hbar\frac{d\rho}{dt} = [H(t),\rho],
\end{equation}
with $\rho$ the density operator of the electron.

\subsection{Electron relaxation}
Eq. \ref{eqSchrodingerMotion} describes the quantum dynamics of an electron with momentum $ \textbf{p} $ that is subject to an external electromagnetic field characterized by its vector potential $ \widetilde{\textbf{A}}(t) $. A valence-band electron can be excited, and become a conduction-band electron by absorbing a photon. The transition happens with a particular probability that depends on the field intensity and photon energy. The excited conduction-band electron could be stimulated into the valence band and emit a photon. This whole process is the Rabi oscillation. However, in a complete physical picture, electrons in the conduction band undergo ultrafast carrier relaxation processes caused by the faster electron-electron scattering followed by the slower electron-phonon scattering \cite{sun08,haug04,winzer10}. These scattering processes destroy the quantum coherence of electrons and result in quantum dephasing. In a strong excitation regime, a strong pump pulse depletes the population in the valence band and excites electrons into the conduction band. Thereafter, the hot electrons in the conduction band undergo an ultrafast electron-electron scattering process (as do the holes in the valence band) that results in quasi-thermalized statistics for electrons in the conduction band and holes in the valence band. The quasi-thermalized distributions establish new state-occupation probabilities within the two bands. As a result, a strong pump not only changes the state-occupation probability of the state having the corresponding energy, but also affects the state-occupation probabilities over the whole band. A complete treatment that would include the electron-photon, electron-electron, and electron-phonon interactions requires sophisticated quantum-mechanical calculations \cite{haug04,kira06}. There may be additional quantum dephasing processes. However, if we limit ourselves to low-excitation quasi-continuous-wave regime, in which the state-occupation probabilities over the whole bands do not change significantly by the pump nor in time, the situation is much simpler. To model the ultrafast quantum dephasing in a low excitation quasi-continuous-wave regime, we simply introduce two phenomenological decay rates $ \Gamma_1 $ and $ \Gamma_2 $, describing the population and quantum-coherence damping that phenomenologically represent the overall effect of the ultrafast scattering processes.  With the two phenomenological decay rates, one arrives at the Bloch equations for graphene under low excitation power:


\begin{eqnarray}
\label{eqSchrodingerTwoLevel}
\dot{\varrho}&=&-\Gamma_1(\varrho-\varrho^{eq}) + \frac{2ei}{\hbar}\sum_{u,j} \widetilde{A}_{u,j}(t)[V_{VC}^u(\rho_{VC}+\rho_{CV})] \notag\\
\dot{\rho}_{VC}&=&\left(-i\omega_{VC}-\Gamma_2\right)\rho_{VC}-\frac{ei}{\hbar}\sum_{u,j} \widetilde{A}_{u,j}(t)[2V_{VV}^u\rho_{VC}+V_{VC}^u\rho] \notag\\
\dot{\rho}_{CV}&=&\left(-i\omega_{CV}-\Gamma_2\right)\rho_{CV}-\frac{ei}{\hbar}\sum_{u,j} \widetilde{A}_{u,j}(t)[-2V_{VV}^u\rho_{CV}+V_{VC}^u\rho],\notag\\
\end{eqnarray}
where $\varrho = \rho_{CC}-\rho_{VV}$ is the population inversion, $ \omega_{CV} = -\omega_{VC} = 2E_C/\hbar $, and $\varrho^{eq}$ is the population inversion in thermal equilibrium. Compared to the Bloch equations for atom vapours and normal semiconductors \cite{haug04,boyd92}, both diagonal and off-diagonal driving terms exist for the off-diagonal elements of the density matrix. General solutions of Eq. \ref{eqSchrodingerTwoLevel} are difficult to find. Instead, we seek perturbative solutions, which are are good approximations when the pump power is well below the saturation threshold, which is measured to be $ 4 $ GW/cm$ ^2 $ \cite{xing10}.



The perturbative method gives the solution in the following cascaded form in the frequency domain, with each order denoted by a positive index $ k $:
\begin{eqnarray}
\label{eqCascadedRhoDecayFreq}
\varrho^{(k+1)}(\Omega)&=&\frac{e}{\hbar}\sum_{u,j}\frac{2A_{u,j}}{(\Omega-i\Gamma_1)}V_{VC}^u\left[\rho^{(k)}_{VC}(\Omega+\omega_{u,j})+\rho^{(k)}_{CV}(\Omega+\omega_{u,j})\right] \notag\\
\rho^{(k+1)}_{VC}(\Omega)&=&- \frac{e}{\hbar}\sum_{u,j} \frac{ A_{u,j}}{(\Omega+\omega_{VC}-i\Gamma_2)}\left[2V_{VV}^u\rho^{(k)}_{VC}(\Omega+\omega_{u,j})+V_{VC}^u\varrho^{(k)}(\Omega+\omega_{u,j})\right] \notag\\
\rho^{(k+1)}_{CV}(\Omega)&=&-\frac{e}{\hbar}\sum_{u,j}\frac{ A_{u,j}}{(\Omega+\omega_{CV}-i\Gamma_2)}\left[-2V_{VV}^u\rho^{(k)}_{CV}(\Omega+\omega_{u,j})+V_{VC}^u\varrho^{(k)}(\Omega+\omega_{u,j})\right],\notag\\
\end{eqnarray}
where $ \Omega = \omega-i\kappa $ with $ \kappa > 0 $ in order for the Laplace transform to converge. 

\section{Surface-current density in graphene}
In this section, the quantum dynamics of electrons will be derived perturbatively. We first derive the first-order solution, which determines the linear optical response of graphene. We then compare our theoretical result with the linear complex conductivity measured by \cite{bruna09,wang08,ni07}. The cascaded relation in Eq. \ref{eqCascadedRhoDecayFreq} allows the derivation of higher-order  density matrix terms, which determine the nonlinear quantum dynamics of electrons in graphene. We continue to calculate up to the third-order terms, which allows the prediction of nonlinear optical responses such as four-wave mixing. We finally fit the theory to recently reported four-wave mixing experimental data \cite{hendry10}.

\subsection{Linear optical conductivity}
\label{secLinearConductivity}
The zeroth-order solution to the density matrix is simply the thermal-equilibrium values of the population inversion and quantum coherence. At temperature $ T = 0 $, we have $\rho^{(0)}_{CC} = \rho^{(0)}_{VC} = \rho^{(0)}_{CV} = 0$, $ \rho^{(0)}_{VV} = 1$. Inserting zeroth-order solution into Eq. \ref{eqCascadedRhoDecayFreq} leads to the first-order solution:
\begin{eqnarray}
\label{eqRho1}
\varrho^{(1)}(\Omega) &=& 0 \notag\\
\rho^{(1)}_{VC}(\Omega) &=& \frac{e}{\hbar}\cdot\frac{1}{(\Omega+\omega_{VC}-i\Gamma_2)}\sum_{w,n} V_{VC}^w \widetilde{A}_{w,n}(\Omega)\notag\\
\rho^{(1)}_{CV}(\Omega) &=& \frac{e}{\hbar}\cdot\frac{1}{(\Omega+\omega_{CV}-i\Gamma_2)}\sum_{w,n} V_{VC}^w \widetilde{A}_{w,n}(\Omega),
\end{eqnarray}
where $\widetilde{A}_{w,n}(\Omega)$ is the Laplace transform of the time-dependent vector potential of one frequency mode indexed by $ n $ on polarization $w$. The expected velocity on $ q $ polarization of a field-coupled electron is given by
\begin{equation}
\langle{v^q}\rangle = Tr\left[v^q\rho\right],
\end{equation}
where the velocity operator $ v^q $ is defined as
\begin{equation}
\label{eqVelocityOperator}
v^q = \frac{\partial H}{\partial p_q} = \left(\begin{array}{cc}
0&\frac{\partial h_{\textbf{p}}}{\partial p_q}\\
\frac{\partial h^*_{\textbf{p}}}{\partial p_q}&0\\
\end{array}\right)
+ e\sum_u \widetilde{A}_u(t)\left(\begin{array}{cc}
0&\frac{\partial^2 h_{\textbf{p}}}{\partial p_u \partial p_q}\\
\frac{\partial^2 h^*_{\textbf{p}}}{\partial p_u \partial p_q}&0\\
\end{array}\right) = v'^q + v''^q.
\end{equation}
To simplify our discussion, we only consider an external field polarized in the $ \hat{x} $ direction with vector potential $ \widetilde{A}(t) $. 


The linear electron velocity is contributed by both the diagonal elements from $ \rho^{(0)} $ and the off-diagonal elements from $ \rho^{(1)} $ \cite{kao10,lewkowicz09}:
\begin{equation}
\langle{v^x}\rangle_l = Tr\left[v'^{\,x}\rho^{(1)}+v''^{\,x}\rho^{(0)}\right].
\end{equation}
The linear surface-current density $ \widetilde{J}_l\left(\Omega\right) $ can be obtained by an integral over the whole Brillouin zone. Since the integrand  $\langle{v^x}\rangle_l $ is only appreciable in the vicinity of the input optical frequency $ \omega $, instead of an integral limited in the Brillioun zone, we can extend the integral to infinity without significantly affecting the result. The integral is written as
\begin{eqnarray}
\label{eqSurfaceCurrentIntegral}
\widetilde{J}_l\left(\Omega\right) &=& \frac{1}{4\pi^2}\int_0^{2\pi}d\theta\int_0^{\infty}-e\langle{v^x}\rangle_l kdk\notag\\
&=& -i\frac{e^2}{16\hbar}(\Omega-i\Gamma_2)\widetilde{A}(\Omega).
\end{eqnarray}
By using the frequency-domain relation $ \widetilde{E}(\Omega) = -i\Omega \widetilde{A}(\Omega) $, the linear optical conductivity is derived as
\begin{equation}
\label{eqSigmaLinear}
\sigma_l = 4\times\frac{e^2}{16\hbar}\left(1+\frac{\Gamma_2}{\omega}i\right) = \sigma_0\left(1+\frac{\Gamma_2}{\omega}i\right),
\end{equation}
where $ \sigma_0 $ is the theoretical prediction of the universal optical conductivity of graphene \cite{nair08,stauber08,mecklenburg10}. A factor $ 4 $ is added due to valley and spin degeneracies. If we compare Eq. \ref{eqSigmaLinear} with the experimental measurements of the complex optical conductivity of graphene at $ \lambda = 550$nm in \cite{bruna09,wang08,ni07}, we find that $ \Gamma_2 $ ranges from $ 1.39 \times 10^{15} $ s$ ^{-1} $ to $ 4 \times 10^{15} $ s$ ^{-1} $ by normalizing the real parts to $ \sigma_0 $.

In \cite{skulason10}, the authors ascribe the imaginary part of the conductivity of graphene to the virtual transition of electrons at M and $ \Gamma $ points in the Brillioun zone. However, our quantum mechanical calculations based on the nearest-neighbour-interaction approximation find that we have $ \nabla h_{\textbf{p}} = 0$ at the $ \Gamma $ point, leading to a zero interaction Hamiltonian, i.e., $ V(t) = 0 $. At the M point,  the interaction Hamiltonian gives pure real diagonal elements, and pure imaginary off-diagonal elements in Eq. \ref{eqVtMatrix}. Thus, the conductivity at the M point is still real as long as we set $ \Gamma_1 = \Gamma_2 = 0 $. Thus it seems that the relatively large imaginary part of the conductivity arises from the ultrafast quantum dephasing in graphene. The fact that this time constant is approximately 1 fs indicates both that optical measurements can be used to characterize quantum dephasing in particular samples and that dipole quantum coherence would seem to be a poor candidate for coherent electron devices in graphene.

\subsection{Nonlinear optical conductivity}
To evaluate the nonlinear interactions in graphene, we need to solve for higher-order density-matrix terms $ \rho^{(2)} $ and $ \rho^{(3)} $. Having obtained $ \rho^{(2)} $ and $ \rho^{(3)} $, the expected nonlinear velocity on polarization $q$ for an electron with momentum $\textbf{p}$ can be calculated as
\begin{equation}
\langle v^{q}\rangle_{nl}= Tr\left[ v^q\rho^{(3)}\right],
\end{equation}
According to Eq. \ref{eqVelocityOperator}, both $ \rho^{(2)} $ and $ \rho^{(3)} $ contribute to $\langle v^{q}\rangle_{nl}$. To derive the nonlinear surface-current density, we need to integrate over the entire Brillouin zone. $ \rho^{(2)} $ and $ \rho^{(3)} $ both are appreciable only in the vicinities of their resonant frequencies. The contribution from $ \rho^{(2)} $ can be neglected since the second derivative of $ h_{\textbf{p}} $ gives zero due to the linear band structure of graphene near the Dirac points. Thus, in the nonlinear surface-current-density calculation, we only keep the $ v'^q $ term and drop the $ v''^q $ term in Eq. \ref{eqVelocityOperator}. The resulting nonlinear velocity is written as
\begin{equation}
\label{eqThirdOrderVelocity}
\langle v^{q}\rangle_{nl} = Tr\left[v'^q\rho^{(3)}\right]=\sum_{u,l}\sum_{v,m}\sum_{w,n} \langle v_{u,l;v,m;w,n}^q\rangle,
\end{equation}
where the expected velocity of one frequency component on polarization $q$ is defined as $\langle v_{u,l;v,m;w,n}^q\rangle$. Since the main contribution of $ \langle v_{u,l;v,m;w,n}^q\rangle $ comes from the vicinities of each resonant frequency, we can extend the integration in the Brillouin zone to infinity and without significantly affecting the result. By assuming the Dirac cone goes to infinity, we obtain analytical solutions to the third-order surface-current density of at one frequency component having polarization $ q $ to be

\begin{eqnarray}
\widetilde{J}_{u,l;v,m;w,n}^q(\Omega) &=& \frac{1}{4\pi^2}\int_0^{2\pi}d\theta\int_0^{\infty}-e\langle v_{u,l;v,m;w,n}^q\rangle kdk\notag\\
&=&-i\frac{v^2_Fe^4}{16\hbar^3}A_{u,l}A_{v,m}\left[\frac{\mu_{q,u,v,w}}{\Omega-i\Gamma_1}+\frac{\nu_{q,u,v,w}(\Omega+\omega_{u,l}+\omega_{v,m}-i\Gamma_2)}{(\Omega+\omega_{u,l}-i\Gamma_1)(2\Omega+\omega_{u,l}+\omega_{v,m}-2i\Gamma_2)}\right] \times \notag \\& & \widetilde{A}_{w,n}(\Omega+\omega_{u,l}+\omega_{v,m}),\notag\\
\end{eqnarray}
where $\mu_{q,u,v,w}$ and $\nu_{q,u,v,w}$ are
\begin{eqnarray}
\mu_{q,u,v,w} = \frac{4}{\pi v^4_F}\int_0^{2\pi}V_{VV}^q V_{VC}^u V_{VV}^v V_{VC}^w d\theta \notag\\
\nu_{q,u,v,w} = \frac{4}{\pi v^4_F}\int_0^{2\pi}V_{VC}^q V_{VC}^u V_{VC}^v V_{VC}^w d\theta .
\end{eqnarray}
In the time domain, the steady-state solution to the nonlinear surface-current density reads
\begin{eqnarray}
\label{eqCurrentDensitySteady}
\widetilde{J}_{u,l;v,m;w,n}^q(t)&=&-4\times\frac{v^2_Fe^4}{16\hbar^3}\frac{E_{u,l} E_{v,m} E_{w,n}}{\omega_{u,l}\omega_{v,m}\omega_{w,n}}\Bigg[\frac{\mu_{q,u,v,w}}{(\omega_{u,l}+\omega_{v,m}+\omega_{w,n}+i\Gamma_1)}\notag\\
&+&\frac{\nu_{q,u,v,w}(\omega_{w,n}+i\Gamma_2)}{(\omega_{u,l}+\omega_{v,m}+2\omega_{w,n}+2i\Gamma_2)(\omega_{v,m}+\omega_{w,n}+i\Gamma_1)}\Bigg]e^{-i(\omega_{u,l}+\omega_{v,m}+\omega_{w,n})t},\notag\\
\end{eqnarray}
where a coefficient $ 4 $ is added to include valley and spin degeneracies. We have substituted the vector-potential amplitude with the electric-field amplitude by use of Eq. \ref{eqEARelation}. Having obtained the surface-current density of one frequency component $\widetilde{J}_{u,l;v,m;w,n}^q(t)$ on polarization $ q $, the total surface-current density on this polarization is the summation over all frequency components,
\begin{equation}
\label{eqJTotal}
\widetilde{J}_{nl}^q(t) = \sum_{u,l}\sum_{v,m}\sum_{w,n}\widetilde{J}_{u,l;v,m;w,n}^q(t).
\end{equation}
Eq. \ref{eqJTotal} is a general expression for the total third-order nonlinear surface-current density, composed of different frequency modes, each of which is produced by a particular nonlinear-optical process. For example, let $\omega_{p_1}, \omega_{p_2}$ be the frequencies of the pump modes and $\omega_s$ be the frequency of the signal mode, the frequency  $ 3\omega_{p_1} $ results from third-harmonic generation of the pump at frequency $ \omega_{p_1} $ while $\omega_i = \omega_{p_1} + \omega_{p_2}-\omega_s$ is produced by four-wave mixing. 

We mark one frequency mode with frequency $\omega$ on polarization $q$ as $\widetilde{J}^q_{\omega}(t)$. $ \widetilde{J}^q_{\omega}(t) $ is the summation over all frequency components $\widetilde{J}_{u,l;v,m;w,n}^q(t)$ satisfying $\omega_{u,l}+\omega_{v,m}+\omega_{w,n}=\omega$:
\begin{equation}
\label{eqCurrentDensity1Mode}
\widetilde{J}_{\omega}^q(t) = \sum_{\omega_{u,l}+\omega_{v,m}+\omega_{w,n}=\omega}\widetilde{J}_{u,l;v,m;w,n}^q(t) = J_{\omega}^q e^{-i\omega t}.
\end{equation}

\section{Four-wave mixing in graphene}
Of particular interest for this paper is the study of four-wave mixing,  a nonlinear-optical process involving four modes.  Two of these serve as the pump modes, one as the signal mode, and one as the idler mode. The optical frequencies $\omega_{p_1}$ and $\omega_{p_2}$ correspond to the two pump modes, $\omega_s$ to the signal mode, and $\omega_i$ to the idler mode. Energy conservation gives $\omega_{p_1}+\omega_{p_2} = \omega_s + \omega_i$. Microscopically, four-wave mixing is a process in which two pump photons are annihilated while one signal and one idler photon are created. In the special case where the two pump modes share the same optical frequency, i.e., $\omega_{p_1} = \omega_{p_2} = \omega_p$, the four-wave mixing is pump degenerate. Since four-wave mixing is a third-order nonlinear process, its strength in the electric field is proportional to the nonlinear susceptibility $\chi^{(3)}$. If the frequency of the optical modes are far away from the resonant frequency of the nonlinear medium,  the results is a non-resonant susceptibility $\chi^{(3)}_{NR}$. In the quantum-mechanical picture, $\chi^{(3)}_{NR}$ results from the fact that the photon energies are far from the bandgap of the nonlinear medium. A typical value for the non-resonant susceptibility in optical fibers is $\chi^{(3)}_{NR}\sim10^{-15}$ esu \cite{boyd92}. However, a much higher $\chi^{(3)}$ \cite{boyd81,nilsen81} is present if the bandgap of the nonlinear medium is close to the interacting-photon energies. The resonance-enhanced nonlinear susceptibility $\chi^{(3)}_R$ is typically of the order of $10^{-7}$ esu.

\begin{figure}
\begin{center}

\includegraphics[scale=0.35]{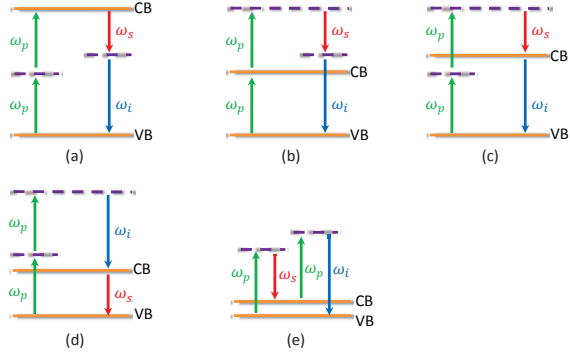}
\caption{\label{Fig:FWMDiagram}Five resonance-enhanced four-wave mixing processes in graphene. (a) Two-photon absorption enhanced four-wave mixing. (b) One-photon absorption enhanced four-wave mixing. (c) Idler resonance-enhanced four-wave mixing. (d) Signal resonance-enhanced four-wave mixing. (e) Detuning-enhanced four-wave mixing.}
\end{center}
\end{figure}

Graphene is a zero-bandgap semiconductor with linear band structure near the Dirac points, making it ``more resonant'' than typical resonant media. The zero bandgap of graphene results in the fact that any frequency within the range from DC to optical frequencies is resonant. This is the physical origin of the high and uniform absorption of graphene. While absorption is a resonance-enhanced linear process, nonlinear processes can also be enhanced by resonance. In particular, four-wave mixing in graphene can be enhanced by the five resonance-enhanced processes plotted in Fig. \ref{Fig:FWMDiagram}. 
Different resonance-enhanced four-wave mixing processes exist for electrons with different energies in graphene. We next discuss these resonance-enhanced four-wave mixing processes.

(a) Two-photon absorption enhanced four-wave mixing:
Two-photon absorption happens when the bandgap of a material is twice the photon energy of the incident light. In graphene, this corresponds to $\omega_{CV} \simeq 2\omega_p$.

(b) One-photon pump absorption enhanced four-wave mixing:
Unlike two-photon absorption, in which one electron is excited by two incoming photons simultaneously, an electron is excited by a single photon in a one-photon absorption process. In graphene, the condition for one-photon absorption is $\omega_{CV} \simeq \omega_p$.  

(c) Idler-enhanced four-wave mixing:
In four-wave mixing, if the emitted light frequency, i.e., $\omega_i$, is close the the bandgap of the nonlinear medium, we also have resonance-enhanced effect. In graphene, this happens at the bandgap $ \omega_{CV} = \omega_i$. \\

(d) Signal-enhanced four-wave mixing:
If the signal-photon energy is close to the bandgap of the nonlinear medium, we also have enhanced effect for four-wave mixing. In graphene, signal-enhanced four-wave mixing happens at bandgap $ \omega_{CV} = \omega_s $. \\

(e) Detuning-enhanced four-wave mixing:
Detuning-enhanced four-wave mixing in graphene is similar to the four-wave mixing enhancement by coherent anti-Stokes Raman scattering \cite{zumbusch99,freudiger08}, when the frequency detuning between the pump mode and the Stokes mode is close to the molecule-vibration energy. In graphene, this is the case when the bandgap is close to the frequency detuning between the pump and the signal, i.e., $\omega_{CV} \simeq \omega_p - \omega_s$. \\

\subsection{Non-degenerate four-wave mixing}
\subsubsection{Co-polarized non-degenerate four wave mixing}
In the co-polarized case, we let all frequency modes be $\hat{x}$ polarized. We define the conductivity of the co-polarized non-degenerate four-wave mixing to be
\begin{equation}
\widetilde{J}_{CN}^x(t) = \sigma_{CN} E_{p_1}E_{p_2}E_s^*e^{-i\omega_i t} + c.c. =\widetilde{J}_{\omega_i}^x(t) + \widetilde{J}_{-\omega_i}^x(t).
\end{equation}
The frequency components contributing to $\sigma_{CN}$ are listed in Tab. \ref{tabCNFWM}.
\begin{table}
\caption{\label{tabCNFWM}Frequency components for co-polarized non-degenerate four-wave mixing}
\begin{center}
\begin{tabular}{c|c|c|c|c}
$\omega_{u,l}$ & $\omega_{v,m}$ & $\omega_{w,n}$ & $\mu_{x,u,v,w}$ & $\nu_{x,u,v,w}$\\
\hline
$x, \omega_{p_1}$ & $x, \omega_{p_2}$ & $x, -\omega_s$ & -1 & 3\\
\hline
$x, \omega_{p_2}$ & $x, \omega_{p_1}$ & $x, -\omega_s$ & -1 & 3\\
\hline
$x, \omega_{p_1}$ & $x, -\omega_s$ & $x, \omega_{p_2}$ & -1 & 3\\
\hline
$x, \omega_{p_2}$ & $x, -\omega_s$ & $x, \omega_{p_1}$ & -1 & 3\\
\hline
$x, -\omega_s$ & $x, \omega_{p_1}$ & $x, \omega_{p_2}$ & -1 & 3\\
\hline
$x, -\omega_s$ & $x, \omega_{p_2}$ & $x, \omega_{p_1}$ & -1 & 3\\
\hline
\end{tabular}
\end{center}
\end{table}
We insert terms in Tab. \ref{tabCNFWM} into Eq. \ref{eqCurrentDensitySteady} and Eq. \ref{eqCurrentDensity1Mode} to obtain
\begin{equation}
\sigma_{CN} = \frac{J_{\omega_i}^x}{E_{p_1}E_{p_2}E_s^*}.
\end{equation}

\subsubsection{Cross-polarized non-degenerate four wave mixing}
\begin{figure}
\begin{center}
\subfigure[Pump detuning from the center $10$THz. $\Gamma_1 = \Gamma_2 = 10^{13}$.]{\label{figChi3_Ratio_G13_PD10}
\includegraphics[scale=0.15]{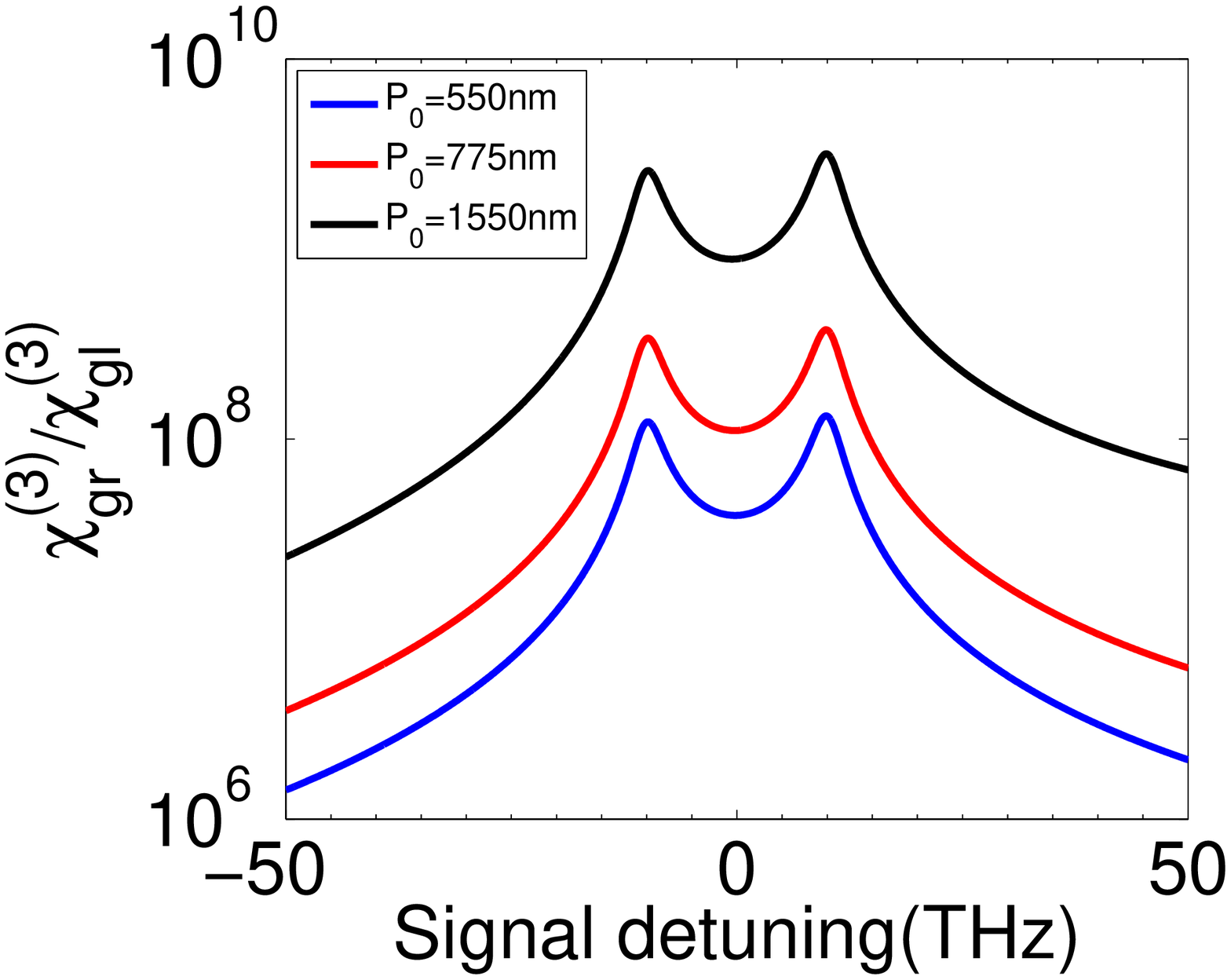}}
\subfigure[Pump detuning from the center $10$THz. $\Gamma_1 = \Gamma_2 = 10^{14}$.]{\label{figChi3_Ratio_G14_PD10}
\includegraphics[scale=0.15]{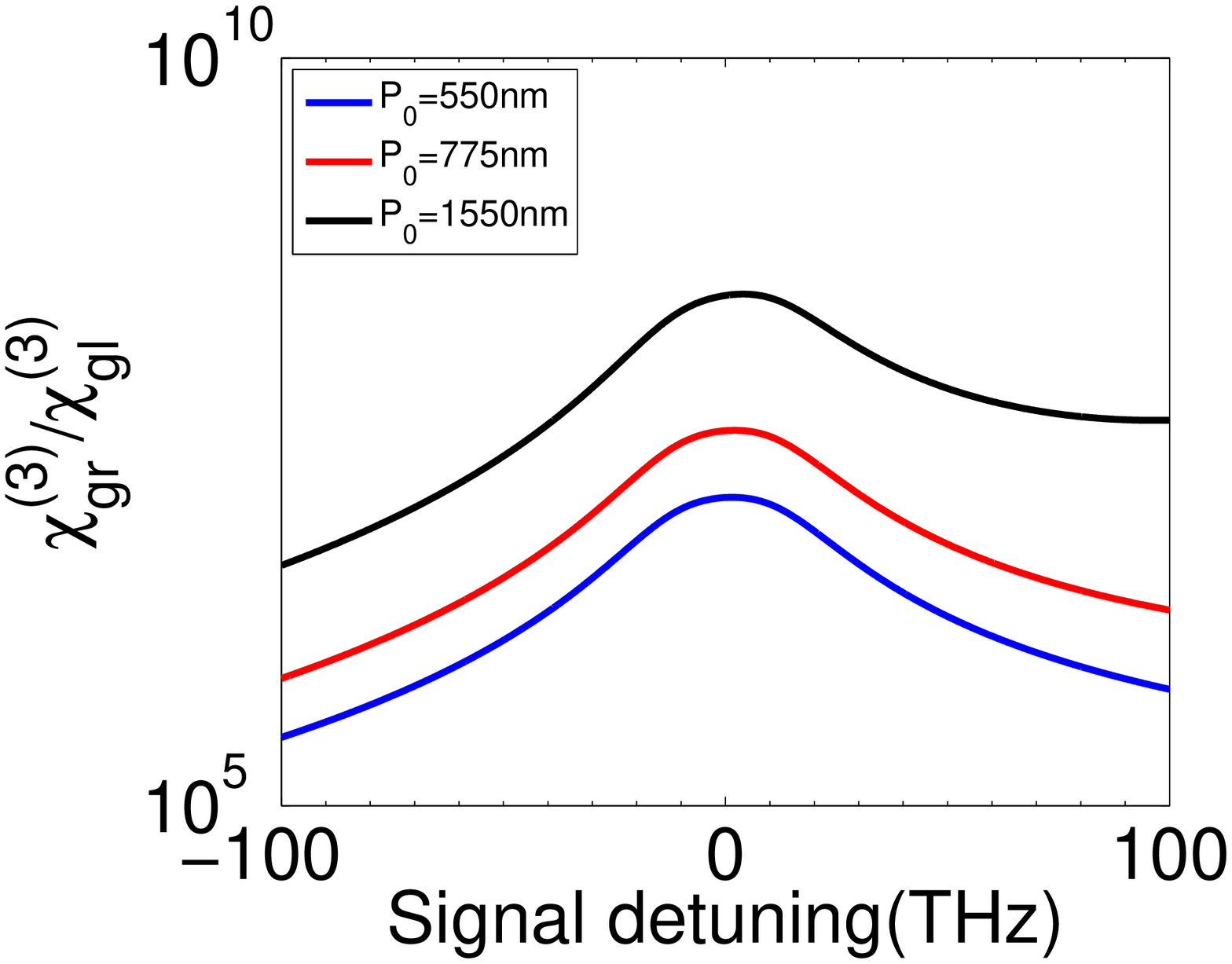}}
\subfigure[Pump detuning from the center $10$THz. $\Gamma_1 = \Gamma_2 = 10^{15}$.]{\label{figChi3_Ratio_G15_PD10}
\includegraphics[scale=0.15]{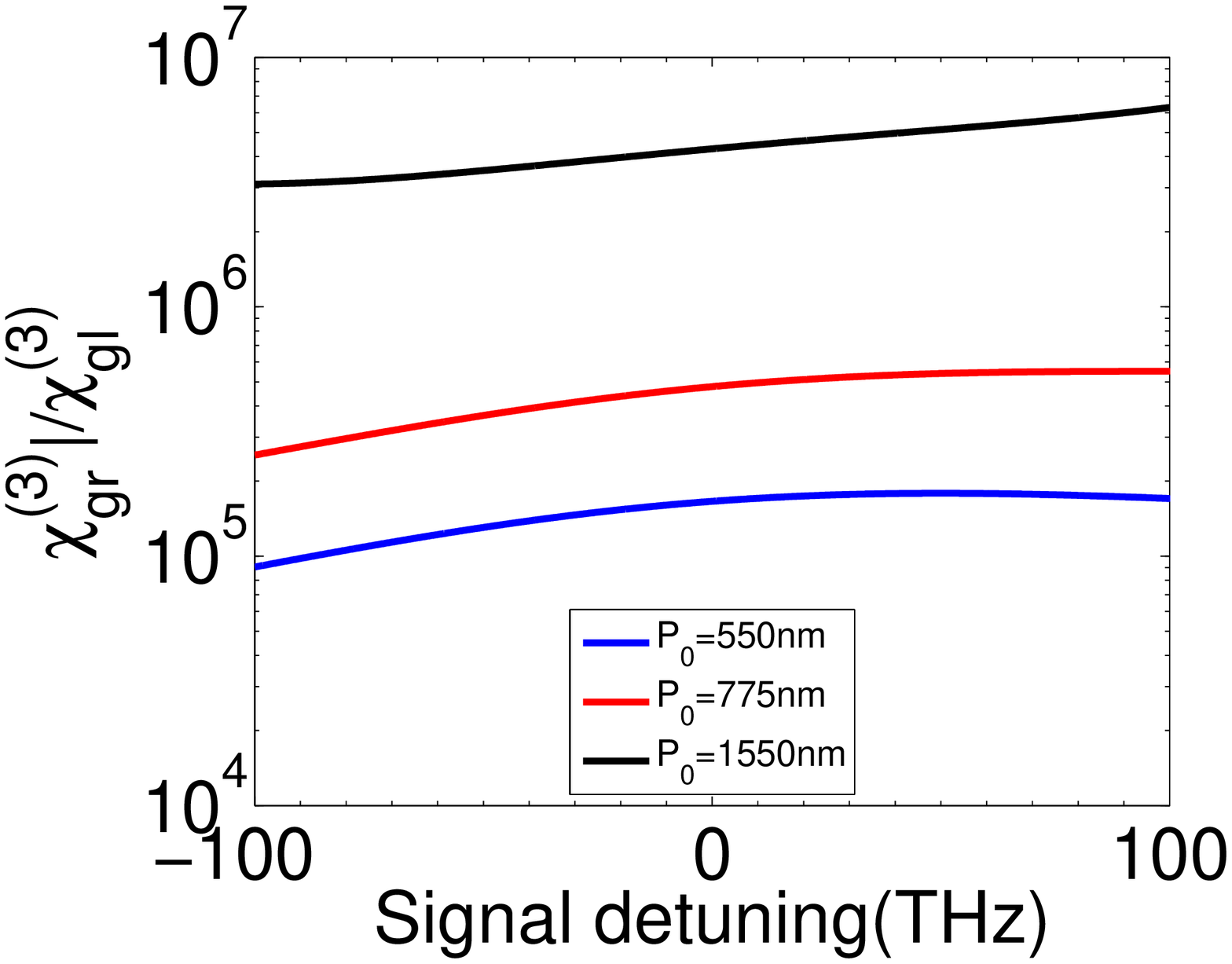}}\\
\subfigure[Pump detuning from the center $20$THz. $\Gamma_1 = \Gamma_2 = 10^{13}$.]{\label{figChi3_Ratio_G13_PD20}
\includegraphics[scale=0.15]{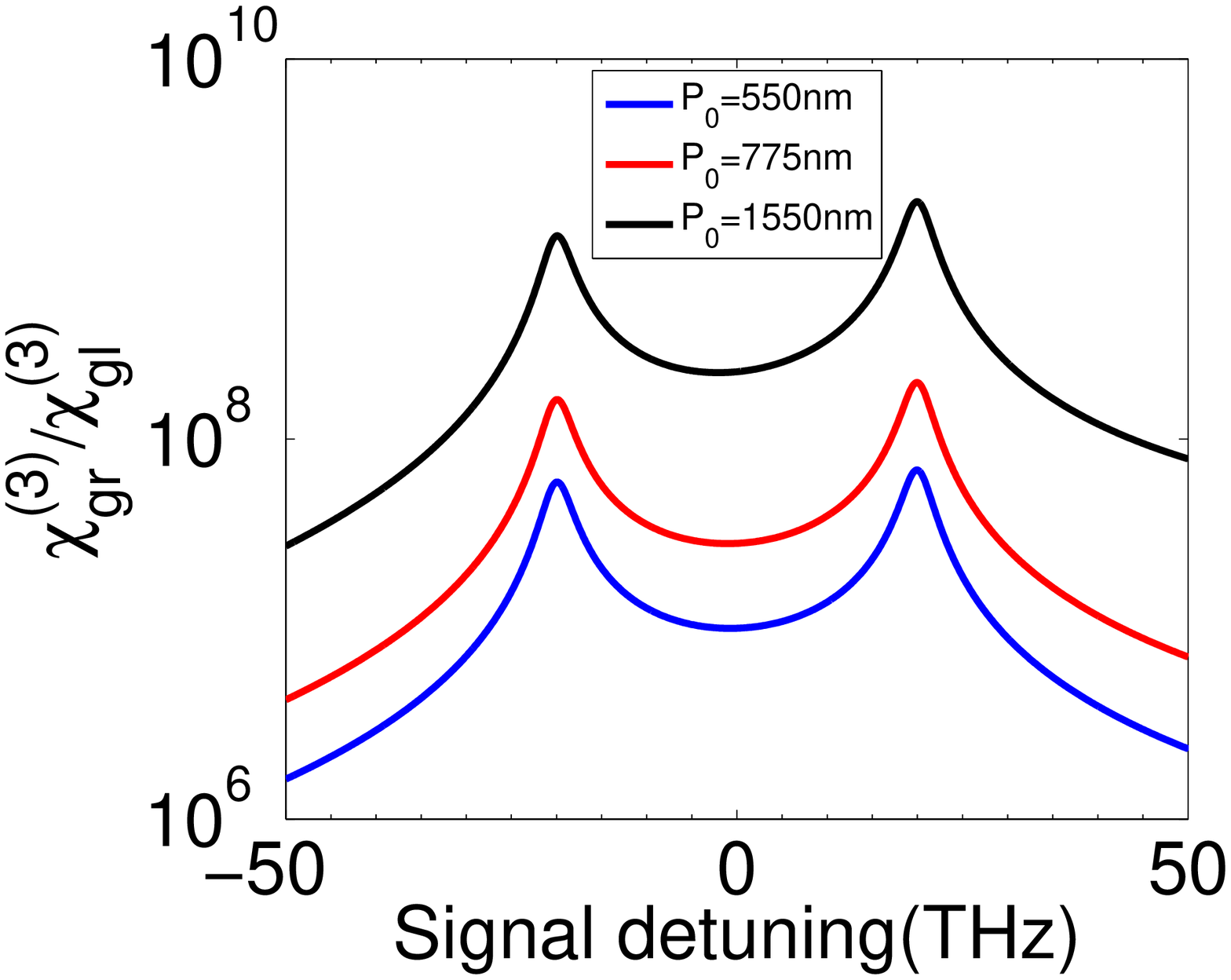}}
\subfigure[Pump detuning from the center $20$THz. $\Gamma_1 = \Gamma_2 = 10^{14}$.]{\label{figChi3_Ratio_G14_PD20}
\includegraphics[scale=0.15]{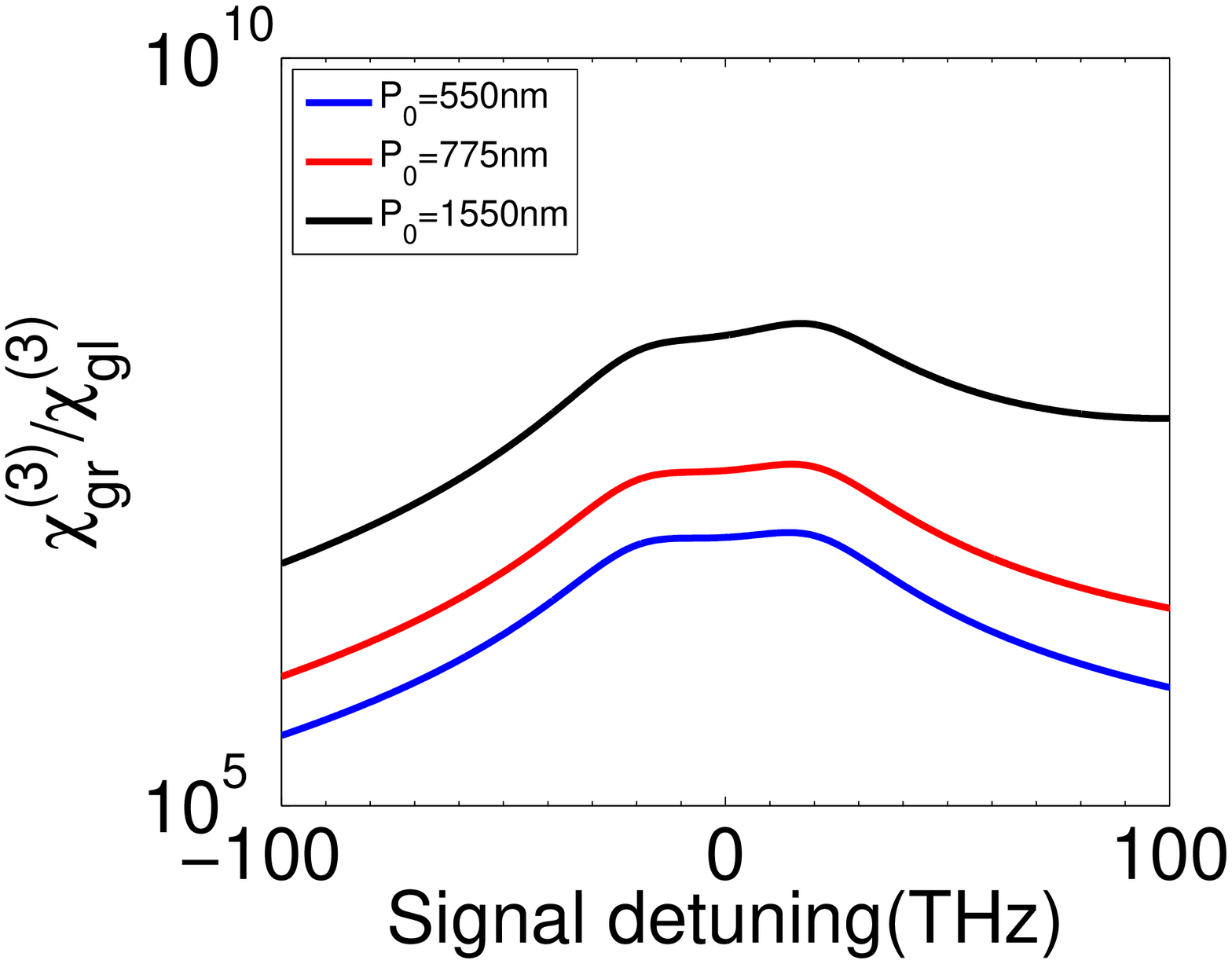}}
\subfigure[Pump detuning from the center $20$THz. $\Gamma_1 = \Gamma_2 = 10^{15}$.]{\label{figChi3_Ratio_G15_PD20}
\includegraphics[scale=0.15]{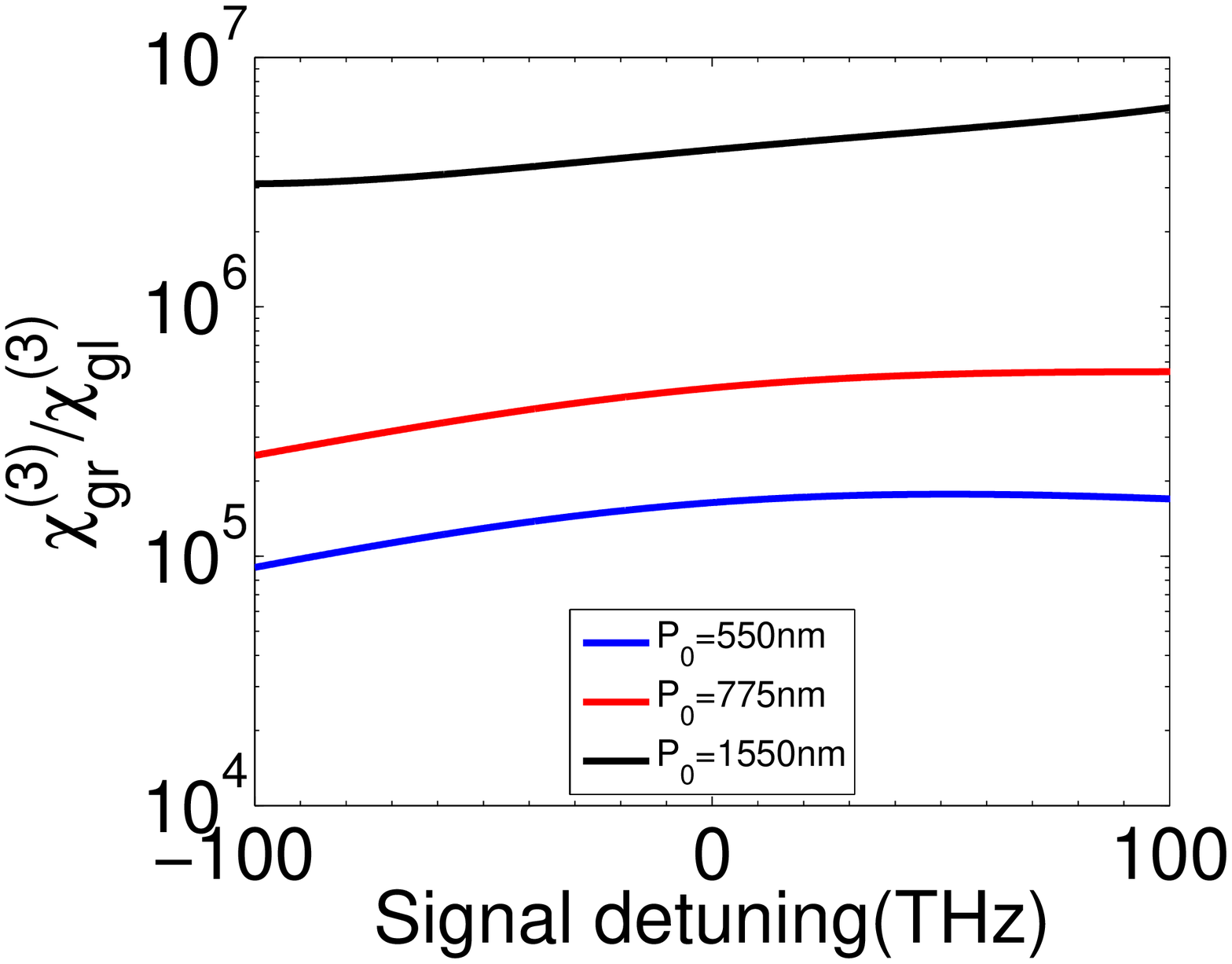}}\\
\subfigure[Pump detuning from the center $30$THz. $\Gamma_1 = \Gamma_2 = 10^{13}$.]{\label{figChi3_Ratio_G13_PD30}
\includegraphics[scale=0.15]{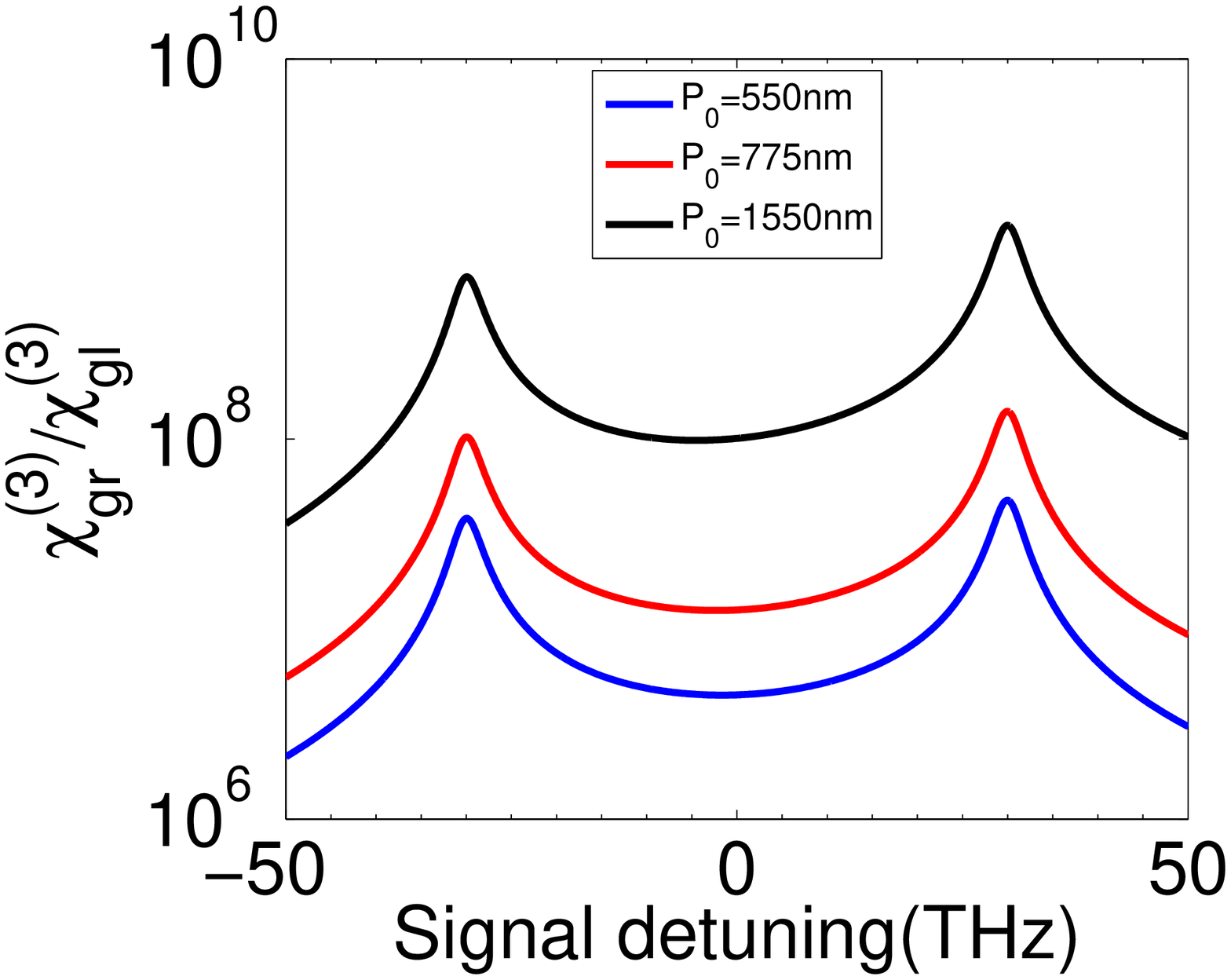}}
\subfigure[Pump detuning from the center $30$THz. $\Gamma_1 = \Gamma_2 = 10^{14}$.]{\label{figChi3_Ratio_G14_PD30}
\includegraphics[scale=0.15]{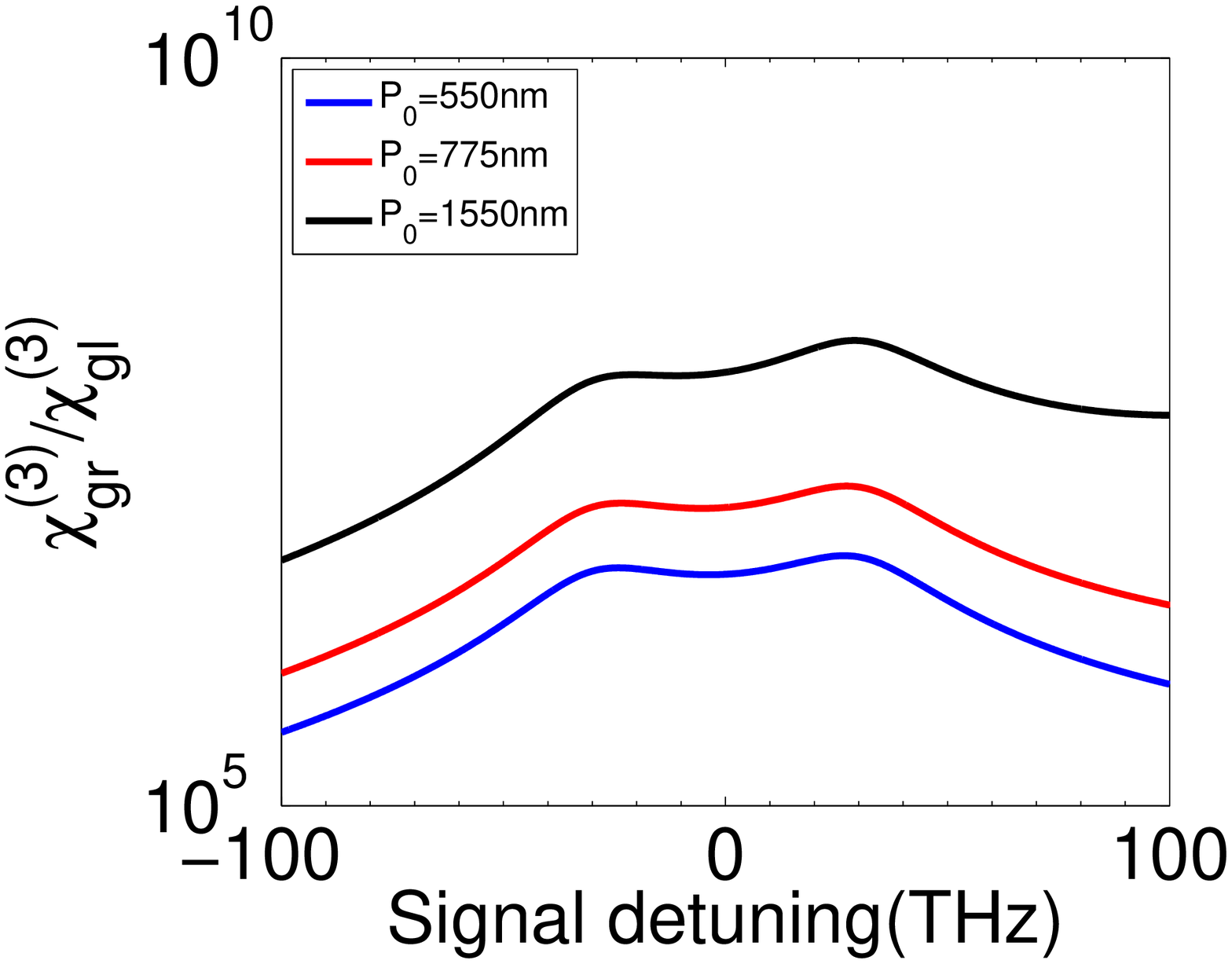}}
\subfigure[Pump detuning from the center $30$THz. $\Gamma_1 = \Gamma_2 = 10^{15}$.]{\label{figChi3_Ratio_G15_PD30}
\includegraphics[scale=0.15]{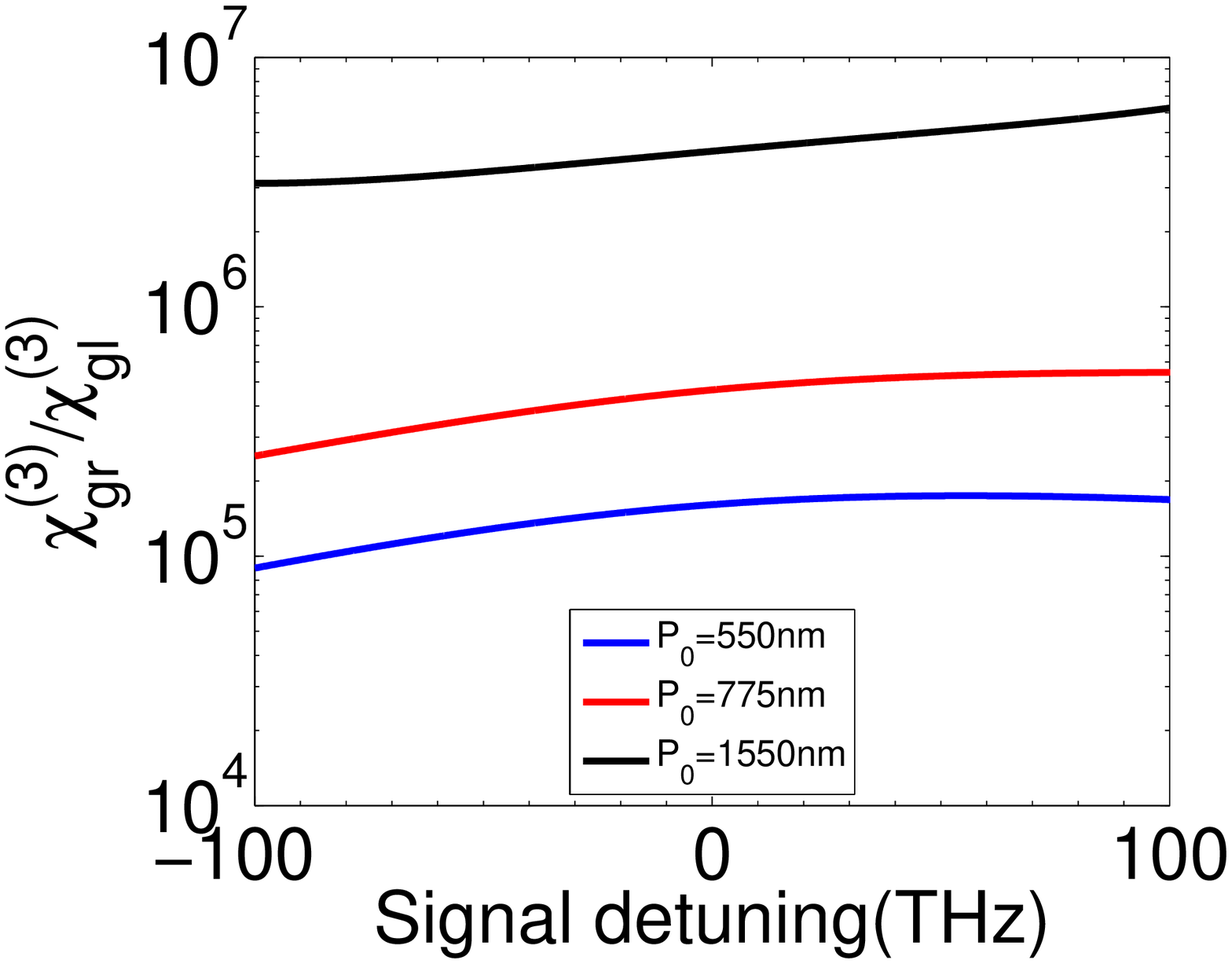}}
\caption{\label{figChi3Ratio} The ratio between the third-order susceptibility of graphene and and that of glass. $ \chi^{(3)}_{gr} $ is calculated from the nonlinear optical conductivity of non-degenerate co-polarized four-wave mixing for different pump detunings and decay rates $\Gamma_1$ and $ \Gamma_2$. $P_0$ denotes the center wavelength between the two pumps.}
\end{center}
\end{figure}

For cross-polarized non-degenerate four-wave mixing, two pumps modes at frequencies $\omega_{p_x}$ and $\omega_{p_y}$ are on orthogonal polarizations $\hat{x}$ and $\hat{y}$. We assume that the signal mode at frequency $\omega_s$ is polarized on $\hat{x}$. Momentum conservation guarantees that the idler mode is produced on the $\hat{y}$ polarization. We define the conductivity for this nonlinear process to be
\begin{equation}
J_{XN}^y(t) = \sigma_{XN}E_{p_x}E_{p_y}E_s^* e^{-i\omega_it}+c.c. = J_{\omega_i}^y(t) + J_{-\omega_i}^y(t).
\end{equation}
The contributed frequency components to $\sigma_{XN}$ are listed in Tab. \ref{tabXNFWM}.
\begin{table}
\caption{\label{tabXNFWM}Frequency components for cross-polarized non-degenerate four-wave mixing}
\begin{center}
\begin{tabular}{c|c|c|c|c}
$\omega_{u,l}$ & $\omega_{v,m}$ & $\omega_{w,n}$ & $\mu_{y,u,v,w}$ & $\nu_{y,u,v,w}$\\
\hline
$x, \omega_{p_x}$ & $y, \omega_{p_y}$ & $x, -\omega_s$ & -3 & 1\\
\hline
$y, \omega_{p_y}$ & $x, \omega_{p_x}$ & $x, -\omega_s$ & 1 & 1\\
\hline
$x, \omega_{p_x}$ & $x, -\omega_s$ & $y, \omega_{p_y}$ & 1 & 1\\
\hline
$x, -\omega_s$ & $x, \omega_{p_x}$ & $y, \omega_{p_y}$ & 1 & 1\\
\hline
$y, \omega_{p_y}$ & $x, -\omega_s$ & $x, \omega_{p_x}$ & 1 & 1\\
\hline
$x, -\omega_s$ & $y, \omega_{p_y}$ & $x, \omega_{p_x}$ & -3 & 1\\
\hline
\end{tabular}
\end{center}
\end{table}
By comparing Table \ref{tabXNFWM} with Table \ref{tabCNFWM}, we find that $\sigma_{XN} = \frac{1}{3}\sigma_{CN}$. Thus, $\sigma_{XN}$ and $\sigma_{CN}$ differ only by a factor of $ 3 $ as in typical nonlinear media. We can adopt the same approach as in \cite{hendry10} to compare the effective third-order susceptibility of graphene with the third-order susceptibility of silica. The effective third-order susceptibility of graphene is given in the MKS units by
\begin{equation}
\chi^{(3)}_{gr} = \frac{|\sigma_{CN}|}{\epsilon_0 \omega_i d},
\end{equation}
where $ d \simeq 3.3 ${\AA} is the effective thickness of a monolayer of graphene. Typical value for non-resonant third-order susceptibility of silica such as glass is $ \chi^{(3)}_{gl} = 2.22\times 10^{-23}$ m$ ^2 $/$ V^2 $ \cite{boyd92}. We plot the ratio between the two susceptibilities in Fig. \ref{figChi3Ratio}. In each figure, three curves correspond to center pump wavelengths at $550$nm, $775$nm and $1550$nm respectively. For each center pump wavelength, we choose three detuning frequencies at $10$THz, $20$THz and $30$THz between the two pump modes.

Although Ref. \cite{hendry10} predicts that the nonlinear susceptibility of graphene is approximately 8 orders of magnitude greater than that of insulators, our more complete model shows that the ratio between the two susceptibilities can vary from 5 to 9 orders of magnitude, depending on the pump wavelengths, the signal detuning, and the decay rates $ \Gamma_1 $ and $ \Gamma_2 $. A faster quantum dephasing results in a lower nonlinearity due to the fact that the coherence of electrons is quickly destroyed by the environment via scattering processes. Our theory also predicts that ss the pump-signal detuning becomes smaller, the strength of four-wave mixing increases due to resonance-enhanced effect. Another interesting phenomenon is that as the quantum dephasing times approach $~$1 fs, the typical symmetry in the nonlinear susceptibility breaks down. 

\subsection{Degenerate four-wave mixing}
\begin{figure}
\begin{center}
\resizebox*{5cm}{!}{\includegraphics{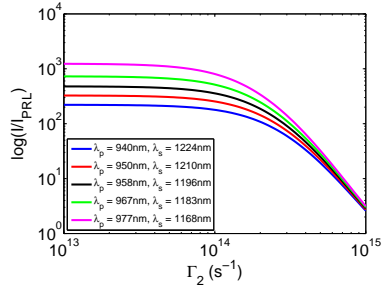}}
\caption{\label{figComparePRL} The four-wave-mixing current intensity ratio between Eq. \ref{eqIidler} and the theoretical result in \cite{hendry10}. Five curves corresponds to five different pump and signal wavelengths in \cite{hendry10}. We set $ \Gamma_1 = 2\Gamma_2 $. Both the ratio and the decay constants are plotted in logarithm scale.}
\label{figFit}
\end{center}
\end{figure}

The optical conductivity for degenerate four-wave mixing can be obtained by setting $ \omega_{p_1}  = \omega_{p_2} = \omega_p$ in the derivation of the non-degenerate four-wave mixing conductivity and dividing the result by $ 2 $ due to degeneracy. In experiment, the observed idler photon-current intensity $ I $ relates to the degenerate four-wave mixing conductivity $ \sigma_D $ by
\begin{equation}
\label{eqIidler}
I  \propto |E_p^2 E_s^* \sigma_D|^2,
\end{equation}
with $ \sigma_D $ as a function of $ \omega_p $, $ \omega_s $, $ \Gamma_1 $ and $ \Gamma_2 $. In Fig. \ref{figComparePRL} we compare the idler photon-current intensity predicted by Eq. \ref{eqIidler} with that is predicted by Eq. 4 in \cite{hendry10}. Depending on the decay rates $ \Gamma_1 $ and $ \Gamma_2 $, the ratio between the two current intensities varies. The result shows that faster quantum-dephasing rates lead to a weaker four-wave mixing. For decay rates at $ 10^{15} $ s$ ^{-1} $, the four-wave-mixing photon-current intensity is roughly equal to what was predicted in \cite{hendry10}. 
We next fit our theory to experimental data obtained by \cite{hendry10}. We use three values for $ \Gamma_2$ found in Sec. \ref{secLinearConductivity} by investigating the imaginary part of the linear complex optical conductivity measured by experiments. Thus, the only parameter to determine is $ \Gamma_1 $. Fits to experimental four-wave-mixing data yields $ \Gamma_1$ ranging from $ 0.8 \times 10^{15} $ s$^{-1} $ to $ 1.25 \times 10^{15} $ s$ ^{-1} $. The fits are shown in Fig. \ref{figFit}. To summarize, the decoherence time is taken from linear refractive index measurement, without fitting. Because the amplitude of the nonlinear response is taken directly from the theory at the decoherence time, only one parameter is used to make the fit.
\begin{figure}
\begin{center}
\resizebox*{5cm}{!}{\includegraphics{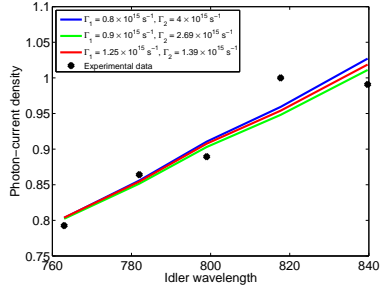}}%
\caption{\label{figFit} Fitting of the theory to the experimental data obtained by \cite{hendry10}. The two phenomenological decay rates for the three curves are $ \Gamma_1 = 0.8 \times 10^{15}$ s$^{-1}, \Gamma_2 = 4 \times 10^{15}$ s$^{-1} $, $\Gamma_1 = 0.9 \times 10^{15}$ s$^{-1}$, $\Gamma_2 = 2.69 \times 10^{15}$ s$^{-1}$, and $\Gamma_1 = 1.25 \times 10^{15}$ s$^{-1}$, $\Gamma_2 = 1.39 \times 10^{15}$ s$^{-1}$.}
\label{figFit}
\end{center}
\end{figure}
The two decay rates were both obtained from experimental data, which provides evidence for ultrafast quantum dephasing in graphene at the 1 fs timescale. Our fully quantum-mechanical model results in faster decay rates than what was reported in \cite{xing10} based on semi-classical calculations. We note here that the decay rates we found are different than carrier-relaxation time \cite{sun08,xing10} in two ways. First, carrier-relaxation time was obtained by differential transmission (DT) experiments where the graphene was excited by an ultrafast strong pump and then probed by a weak pulse. In the probing phase, only electron-electron and electron-phonon interactions exist. However, in a continuous-wave or quasi-continuous-wave experiment, the relatively strong pump (still weaker than the saturation threshold) interact with the graphene through the whole process. Thus, we expect the decay rates to be faster since electron-photon, electron-electron and electron-phonon interactions all happen simultaneously. Second, DT experiments measure the population inversion at the given time. However, our decay rates also include the damping of quantum coherence of electrons. In addition, pure dephasing processes may exist in graphene.

\section{Discussions}
Finding a correct Hamiltonian for a quantum system is of course the key step in derive its dynamics. As we have discussed, we derive the Hamiltonian of electrons in graphene by the minimal substitution $ \textbf{p} \rightarrow \textbf{p} - q\textbf{A} $. The Hamiltonian obtained by minimal substitution is general and it produces the correct electron-photon interactions. However, sometimes the dipole interaction Hamiltonian $ H_I = -e \textbf{r} \cdot \textbf{E} $ was used for semiconductors \cite{haug04} and atom vapours \cite{boyd81}. The dipole interaction Hamiltonian can be derived from the original electron Hamiltonian by the minimal substitution to get
\begin{equation}
\label{eqHamiltonianNormal}
H = \frac{(\textbf{p}-q\textbf{A})^2}{2m}+V(\textbf{r}),
\end{equation}
followed by replacement of $ \textbf{p} $ by means of the commutation relation \cite{haug04}
\begin{equation}
\label{eqSubPtor}
[\textbf{r},H_0] = \frac{i\hbar\textbf{p}}{m}.
\end{equation}
We note here that Eq. \ref{eqSubPtor} only holds for parabolic energy-dispersion relations. However, the energy dispersion is linear near the Dirac points in graphene. Thus, the dipole interaction Hamiltonian does not apply to graphene in the area near the Dirac points. Near the M point and $ \Gamma $ point in graphene, we do have parabolic energy-dispersion relations. If we compare the graphene Hamiltonian in Eq. \ref{eqGrapheneHamiltonian} with the typical Hamiltonian in Eq. \ref{eqHamiltonianNormal} for normal materials, we find that the Hamiltonian of graphene commutes with the momentum operator, and therefore we have energy-momentum degeneracy in graphene. At the M point and the $ \Gamma $ point, the transition matrix element $\langle C_\textbf{p}|\textbf{p}| V_\textbf{p}\rangle = 0$. Thus, as previously discussed, electrons near the $ \Gamma $ point do not interact with the field in the tight-binding model within the nearest neighbour interaction approximation. Electrons near the M point only contribute to the real part of the conductivity if the quantum-dephasing is not present.

The perturbative approach is a good approximation when the saturation effect can be neglected. When saturation effect starts to become significant, the population inversion $\varrho$ becomes so large that our perturbative solution can fail \cite{reichardt96}. However, as we have shown, there exist five different resonance-enhanced four-wave mixing processes in graphene. We find that the population inversion $\varrho^{(2)}$ approaches to its maximal value around $\omega_{CV} \simeq \omega_p$, induced by the one-photon absorption. Thus, the one-photon absorption enhanced four-wave mixing is the most likely to saturate. If we inspect the purturbative solution $ \varrho^{(2)} $, we find that population inversion induced by one-photon absorption is a second-order process. In comparison, the population inversions induced by two-photon absorption, pump-signal detuning, or idler simulation are third-order processes with much weaker magnitudes. Although the signal mode can also induce second-order population inversion, it can be neglected when the strong pump approximation is valid, since the signal intensity is much less than the pump intensity. Under high excitation power levels that can completely change the state-occupation probabilities over the valence and the conduction band, a full-quantum-mechanical many-body theory is required to derive the quantum dynamics of graphene \cite{haug04,kira06}. The quantum many-body theory can also be used to calculate the instantaneous response of electrons in graphene. For steady-state electrons under high excitation power, a quasi-thermalized carrier statistics is established within both the valence and the conduction band. Knowing this quasi-thermalized carrier statistics, one could plug this into $ \rho^{eq} $ in Eq. \ref{eqSchrodingerTwoLevel} and derive the steady-state quantum-dynamics of electrons under high excitation power. 

In our calculations, we made a linear approximation of the band structure of graphene near the Dirac points. This approximation is valid up to $ 4 $eV of bandgap given the hopping energy $ \eta \approx 3 $ eV. For photon wavelengths below $ 500 $nm, the approximated linear band structure starts to be inaccurate and the optical conductivity diverges from the universal value $ \sigma_0 $. A full-band calculation \cite{klintenberg09} would be needed for this situation.

\section{Conclusions}
We have used a quantum-mechanical model to investigate the quantum dynamics of electrons in graphene. The tight-biding model with nearest neighbour approximation and linear band structure near the Dirac points are assumed in order to obtain analytical results. We have also included ultrafast decay and quantum-dephasing effects by adding two phenomenological decay constants into the model. Matches to linear complex optical conductivity and fits to nonlinear four-wave mixing experiments establish quantum-dephasing rates of approximately one femtosecond. This provides evidence for ultrafast dipole quantum-dephasing in graphene. Our approach shows that continuous wave or quasi-continuous-wave optical measurements can yield information about ultrafast scattering processes in graphene. This work also has relevance for proposed coherent electronics applications.

\section*{Acknowledgments}
We gratefully acknowledge support from NSF Grant DMR-0820382, the Embassy of France PUF. We also acknowledge useful discussions with Prof. B. Rosenstein and Prof. D. Citrin.

\end{document}